\documentclass[a4 paper, 11pt,twoside]{article}
\usepackage{fullpage}                           
\usepackage[lmargin=0.6in,rmargin=0.6in,tmargin=0.6in,headsep=.2in]{geometry}
\usepackage{graphicx}
\usepackage{forloop}
\usepackage[dvipsnames]{xcolor}
\usepackage[normalem]{ulem}
\usepackage{dsfont}
\usepackage{fancyhdr}
\usepackage{amsmath,amsfonts,amssymb,amsthm,epsfig,epstopdf,url,array}
 \usepackage[retainorgcmds]{IEEEtrantools}
 \usepackage{makeidx,epsfig,lscape}
 \usepackage{xcolor,pict2e}
\usepackage{upgreek}

 \newtheorem{theorem}{Theorem}[section]
 
 \newtheorem{proposition}[theorem]{Proposition}
 
 \newtheorem{remark}[theorem]{Remark}
 
 \DeclareMathOperator*{\argmax}{argmax}

 \def\te{\theta}
 \def\g{g}
 \def\f{f}
 \def\E{\mathrm{E}}
 \def\R{\mathbb{R}}
 \def\obs{y}
 \def\mis{z}
 \def\OBS{Y}
 \def\obs{y}
 \def\MIS{Z}
 \def\nbind{n}
 \def\nobs{J}

 \def\fisher{I}
 \def\Fisher{I}
 \def\estiun{I_{\nbind,sco}}
 \def\estideux{I_{\nbind,obs}}
 \def\estiemp{I_{\nbind,cov}}

 \newcommand{\Rset}[1]{\mbox{$\mathbb{R}^{#1}$}}
 \newcommand{\Nset}{\mbox{$\mathbb{N}$}}
 
 \newcommand{\Sr}{\mbox{$\mathcal{S}$}}
 \newcommand{\Ir}{\mbox{$\mathcal{I}$}}
 
 \newcommand{\htheta}{{\widehat{\theta}}}

 \newcommand{\pscal}[2] { \left< #1 , #2 \right> }

 \newcommand{\Dt}[1]{\partial_\theta #1}

 \newcommand{\Fr}{\mbox{$\mathcal{F}$}}
 \newcommand{\Lr}{\mbox{$\mathcal{L}$}}

 \def\vectestiun{\bar{I}_{\nbind,sco}}

\begin{document}
{\noindent\huge\bf Estimating Fisher Information Matrix in Latent Variable Models based on the Score Function }\\[0.3cm]

\noindent {\bf\large Maud Delattre}\\[0.3cm]
Universit\'e Paris-Saclay, INRAE, MaIAGE, 78350, Jouy-en-Josas, France\\
Email: maud.delattre@inrae.fr\\[0.3cm]
{\bf\large Estelle Kuhn}\\[0.3cm]
Universit\'e Paris-Saclay, INRAE, MaIAGE, 78350, Jouy-en-Josas, France\\
Email: estelle.kuhn@inrae.fr
\vspace{0.5cm}\\
\noindent {\bf{\large Abstract}}\\
The Fisher information matrix (FIM) is a key quantity in statistics. However its exact computation is often not trivial. In particular in many latent variable models, it is intricated due to the presence of unobserved variables. Several methods have been proposed to approximate the  FIM when it can not be evaluated analytically.  Different  estimates have been considered, in particular moment estimates. However some of them require to compute second derivatives of the complete data log-likelihood which leads to some disadvantages. In this paper, we focus on the empirical Fisher information matrix defined as an empirical estimate of the covariance matrix of the score, which only requires to compute the first derivatives of the log-likelihood. Our contribution consists in presenting a new numerical method to evaluate this empirical Fisher information matrix in latent variable model when the proposed estimate can not be directly analytically evaluated. We propose a stochastic approximation estimation algorithm to compute this estimate as a by-product of the parameter estimate. We evaluate the finite sample size properties of the proposed estimate and the convergence properties of the estimation algorithm through simulation studies. 
\vspace{0.5cm}

\section{Introduction}

The Fisher information matrix (FIM) is a key quantity in statistics as it is required for examples  for evaluating asymptotic precisions of parameter estimates, for building  optimality criteria in experimental designs, for computing Wald test statistics  or classical asymptotic distributions in statistical testing \cite{VanderVaart2000}. It also appears more recently in post model selection inference \cite{charkhi2018asymptotic}, in asymptotic distribution of the likelihood ratio test statistics when testing variance component in mixed models \cite{baey2019asymptotic}   or  as a particular Riemannian metric on complex manifold \cite{le2021fisher}. However its exact computation is often not trivial. This is in particular the case in many latent variables models,  also called incomplete data models, due to the presence of the unobserved variables. Though these models are increasingly used in many fields of application, such as in ecophysiology \cite{Technow2015},   in genomic \cite{Picard2007} or in ecology \cite{Gloaguen2014}. They especially allow a better consideration of the different variability sources and when appropriate, a more precise characterization of the known mechanisms at the origin of the data. When the FIM can not be exactly computed, people either approximate it numerically, for example by using Monte Carlo technics like developed in the R package MIXFIM \cite{mixfim2018} or focus on an estimate of the FIM. The probably most widely used  is the observed FIM \cite{efron1978assessing}.  When  it can not be directly computed in latent variable models, several methods have  been proposed to approximate it. Among the most frequently used approaches are Monte-Carlo methods or iterative algorithms derived from the missing information principle \cite{Woodbury1972}. Indeed according to this principle, the observed Fisher information matrix can be expressed as the difference between two matrices corresponding to the complete information and the missing information due to the unobserved variables (see \textit{e.g.} \cite{McLachlan2008} chapter 4). It enables the development of alternative methods to compute the observed FIM: the Louis's method \cite{Louis1982},  combined with a Monte Carlo method or a stochastic approximation  algorithm by \cite{Delyon1999}, the Oakes method \cite{Oakes1999} or the supplemented Expectation Maximization algorithm \cite{Meng1991}.  However as the observed FIM involves the second derivatives of the observed log-likelihood, all these methods require to compute second derivatives of the complete data log-likelihood which leads to some disadvantages from a computational point of view. More recently, \cite{Meng2017} proposed an accelerated algorithm based on numerical first order derivatives of the conditional expectation of the log-likelihood. Another estimate is the empirical Fisher information matrix. This estimator of the FIM is defined as the moment estimate of the covariance matrix of the score. It is  much less used  than the observed Fisher information matrix. However it has a nice property since it is positive definite, which is not systematically the case for the latter and  it is numerically more interesting because it  only requires the calculation of the first derivatives of the log-likelihood.

In this paper, our contribution consists in presenting a new numerical method to evaluate the empirical FIM in latent variables model. 
Indeed, when the proposed estimate can not be directly analytically evaluated, we propose a stochastic approximation estimation algorithm to compute  it, which  provides this estimate of the FIM as a by-product of model parameter estimates.

The paper is organized as follows. In Section 2, we recall the three main FIM estimates  and discuss their immediate properties. In Section 3, we give practical tools for the computation of the empirical  Fisher information matrix in incomplete data models. In particular, we introduce a new stochastic approximation procedure based on the first derivatives of the complete log-likelihood only and state its asymptotic properties. In Section 4, we illustrate the finite sample size properties of both estimators and the convergence properties of the computation algorithm through simulations. The paper ends by a discussion.

\section{Moment estimates of the Fisher information matrix}

Let us consider a random variable $\OBS$.  Assume $\OBS$ admits a density $g$ with respect to a given measure depending on some parameter $\theta$ taking values in an open subset $\Theta$ of $\Rset{d}$, such that the log-likelihood function $\log g$ is differentiable on $\Theta$ and that $\|\partial_\te  \log g(\obs;\te) (\partial_\te  \log g(\obs;\te))^t\|$ is integrable, where $x^t$ stands for the transpose of a vector or a matrix $x$. Then, by definition, the Fisher information matrix is given for all $\te\in\Theta$ by:
\begin{equation}
	\label{eq:fisher_der1}
	\Fisher(\te) =  E_\te\left[\partial_\te \log g(\OBS;\te) (\partial_\te \log g(\OBS;\te))^t \right].
\end{equation}
When this expression can not be analytically evaluated, people are interested in computing an estimate of the Fisher information matrix.
Considering this expression, one can derive a first moment estimator of the Fisher information matrix based on a $n$-sample $(\obs_1, \ldots, \obs_{\nbind})$ of observations:
	\begin{eqnarray*} 
		\label{estiun}
		\estiun(\te)&=& \frac{1}{n} \sum_{i=1}^n \partial_\te \log \g(\obs_i;\te) (\partial_\te \log \g(\obs_i;\te))^t.
	\end{eqnarray*}
	This estimate is indeed equal to the mean of the Gram matrices of the scores.
	One can also derive a second estimate from (\ref{eq:fisher_der1})  defined as
	\begin{equation*}
		\estiemp(\te) = \frac{1}{n} \sum_{i=1}^n \partial_\te \log \g(\obs_i;\te) (\partial_\te \log \g(\obs_i;\te))^t-\bar{s}\bar{s}^t,
	\end{equation*}	
	where $\bar{s}=\frac{1}{n}\sum_{i=1}^n \partial_\te \log \g(\obs_i;\te)$ (see \textit{e.g.}  \cite{Scott2002}).	We emphasize here that the terminology "empirical Fisher information matrix" is used  in the literature for both estimates (see \textit{e.g.} \cite{kunstner2019limitations}).

If the log-likelihood function $\log g$ is twice differentiable on $\Theta$, the following relation also holds for all $\theta \in \Theta$:
\begin{equation}
	\label{eq:fisher_der2}
	\Fisher(\te) =  - E_\te\left[\partial_\te^2 \log g(\OBS;\te) \right].
\end{equation}

Considering this second expression, we can derive another moment estimator of  the Fisher information matrix based on a $n$-sample $(\obs_1, \ldots, \obs_{\nbind})$ of observations, called the observed Fisher  information matrix defined as:
\begin{eqnarray*} 
	\label{estideux}
	\estideux(\te)&=& - \frac{1}{n} \sum_{i=1}^n \partial_\te^2 \log \g(\obs_i;\te).
\end{eqnarray*}

	\begin{remark}
		We emphasize that the estimate $\estiun(\te)$ is always positive definite, since it is a mean of Gram matrices, contrary to the others estimates $\estideux(\te)$ and $\estiemp(\te)$.
	\end{remark}

	\begin{remark}
		The asymptotical properties of the estimates $\estiun(\te)$ and $\estideux(\te)$ are straighforward when considering independent and identically distributed sample $(\obs_1, \ldots, \obs_{\nbind})$. In particular, assuming standard regularity conditions on $\g$, it follows directly from the  central limit theorem that $\estiun(\te)$ and $\estideux(\te)$ are asymptotically normal. 
	\end{remark}

	\begin{remark}Since both estimators  $\estiun(\te)$ and $\estideux(\te)$ are moment estimates of $\Fisher(\te)$, they are unbiased for all $\te \in \Theta$. This is not the case for $\estiemp(\te)$. Regarding the variance, none of both estimators is better than the other one. This can be highlighted through the following examples. First consider a Gaussian sample with unknown expectation and fixed variance. Then, the variance of the estimator $\estideux(\te)$ is zero whereas the variance of the estimator $\estiun(\te)$ is positive. Second consider a centered Gaussian sample with unknown variance. Then, the variance of $\estiun(\te)$ is smaller than the variance of $\estideux(\te)$. Therefore, none of both estimators is more suitable than the other in general from this point of view. 
	\end{remark}

If the variables $\OBS_1, \ldots, \OBS_{\nbind}$ are not identically distributed, for example if they depend on some individual covariates which is often the case, we state the following result under the less restrictive assumption of  independent non identically distributed random variables.

\begin{proposition}
	Assume that $\OBS_1, \ldots, \OBS_{\nbind}$ are independent non identically distributed random variables each having a parametric probability density function $\g_i$ depending on some common parameter $\te $ in  an open subset $\Theta$ of $\R^p$. Assume also that for all $i$ the function $\log \g_i$ is differentiable in $\te$ on $\Theta$ and that for all $\te \in \Theta$, $\partial_\te \log \g_i(\obs;\te) (\partial_\te \log \g_i(\obs;\te))^t$ is integrable. Moreover assume that for all $\theta$   in $\Theta$, $\lim \frac1{n} \displaystyle\sum_{i=1}^n E_\te(\partial_\te \log \g_i(\obs;\te) (\partial_\te \log \g_i(\obs;\te))^t)$ exists and denote it by $\nu(\theta)$.
	Then, for all $\te \in \Theta$, the  estimator $\estiun(\te)$ is defined, converges almost surely toward $\nu(\theta)$  and is asymptotically normal. 
	Assuming additionaly that $\log\g_i$ is twice differentiable in $\te$ on $\Theta$, the   estimator $\estideux(\te)$ is defined, converges almost surely toward $\nu(\theta)$  and is  asymptotically normal.
\end{proposition}

\begin{proof}
	We prove the consistency by applying the law of large numbers for non identically distributed variables. We establish the normality  result by using characteristic functions. By recentering the terms $E_\te(\partial_\te \log \g_i(\obs;\te) (\partial_\te \log \g_i(\obs;\te))^t)$, we can assume that $\nu(\theta)$ equals zero. Let us denote by $\phi_Z$ the characteristic function for a random variable $Z$. We have for all real $t  $ in a neighborhood of zero that:
	\begin{eqnarray*}
		\| \phi_{\estiun(\theta)/ \sqrt{n}}(t)-1 \|  &=& \|  \prod_{i=1}^n (\phi_{\partial_\te \log \g_i(\obs;\te) (\partial_\te \log \g_i(\obs;\te))^t}(t/n)-1) \| \\
		& \leq &  \prod_{i=1}^n \|  \phi_{\partial_\te \log \g_i(\obs;\te) (\partial_\te \log \g_i(\obs;\te))^t}(t/n)-1 \| 
	\end{eqnarray*}
	Computing a limited expansion in $t$ around zero, we get the result.

	Noting that for all $1 \leq  i \leq \nbind$, $E_\te(\partial_\te \log \g_i(\obs;\te) (\partial_\te \log \g_i(\obs;\te))^t)=-E_\te(\partial_\te^2 \log \g_i(\obs;\te))$, we get the corresponding results for the estimator $\estideux(\te)$.
\end{proof}

\begin{remark}
	The additional assumptions required when considering non identically distributed random variables
	are in the same spirit as  those usually used in the literature. Let us quote for example \cite{Nie2006},  \cite{Silvapulle2011}, \cite{baey2019asymptotic}.
\end{remark}

\section{Computing the  estimator $\estiun(\te)$ in latent variable model}

Let us consider independent random variables $\OBS_1, \ldots, \OBS_{\nbind}$. Assume in the sequel that there exist independent random variables $\MIS_1, \ldots, \MIS_{\nbind}$ such that for each $1 \leq  i \leq  \nbind$, the random vector  $(\OBS_i,\MIS_i)$ admits a   parametric probability density function denoted by $f$ parametrized by $\theta \in \Theta$. We present in this section dedicated tools to compute the estimator $\estiun(\te)$ in latent variable model when it can not be evaluated analytically.

\subsection{Analytical expressions in latent variable models}

In  latent variable models, the estimator $\estiun(\te)$ can be expressed using the conditional expectation as  stated in the following proposition.
\begin{proposition}
	\label{prop:fisherequality}
	Assume that for all $\te \in\Theta$ the function $\log \g(\cdot;\te)$ is integrable, 
	that for all $y$ the function $\log \g(y;\cdot)$ is   differentiable on $\Theta$
	and  that there exists an integrable function $ m$ such that for all  $\te \in\Theta$,   $\|\ \partial_\te \log \g(\obs;\te) \| \leq m(\obs)$. Then for all $\te \in \Theta$ and all $n \in \mathbb{N}^*$:
	\begin{eqnarray*}
		\estiun(\te) &=& \frac{1}{n} \sum_{i=1}^n     \E_{\MIS_i|\OBS_i;\te} (\partial_\te \log \f(\OBS_i,\MIS_i;\te) ) \E_{\MIS_i|\OBS_i;\te} (\partial_\te \log \f(\OBS_i,\MIS_i;\te) ) ^t, 
		\label{eq:vnmis}
	\end{eqnarray*}
	where $\E_{\MIS|\OBS;\te}$ denotes the expectation under the law of $\MIS$ conditionally to $\OBS$.
\end{proposition}
We apply the classical Fisher identity \cite{fisher1925} to  establish the equality stated in Proposition \ref{prop:fisherequality}.
This statement  is indeed in the same spirit  as the well-known Louis formulae for the observed Fisher information matrix estimate \cite{Louis1982}.

\begin{remark}
	In some specific cases  the conditional expectations involved in the previous proposition  admit exact analytical expressions for example  in mixture models which are  developed in  Section \ref{sec:simul} in some simulation studies. 
\end{remark}

\subsection{Computing $\estiun(\te)$ using stochastic approximation algorithm}

When exact computation of the estimator $\estiun(\te)$ is not possible, we propose  to evaluate its value by using a stochastic algorithm which provides the estimate $\estiun(\te)$ as a by-product of the parameter estimates of $\te$.  

\subsubsection{Description of the algorithm in curved exponential family model}
We consider an extension of the stochastic approximation Expectation Maximization algorithm proposed by \cite{Delyon1999} which allows to compute the maximum likelihood estimate in general latent variables model. We assume that the complete log-likelihood belongs to the curved exponential family  for stating the theoretical results. However our algorithm can be easily extended to general latent variables models (see Section \ref{subseq:generalmodel}). As our estimate involves individual conditional expectations, we have to consider an extended form of sufficient  statistics for the model.  Therefore we  introduce the following notations and assumptions.

The individual complete data likelihood function is given for all $1 \leq  i \leq  \nbind$  by:
\begin{equation*} 
	f_i(z_i;\theta)
	= \exp\left(-\psi_i(\theta) + \pscal{S_i(z_i)}{\phi_i(\theta)}\right),
	\label{eq:curvedexpo}
\end{equation*}
where  $\pscal{\cdot}{\cdot}$ denotes the scalar product, $S_i$ is a 
function on $\mathbb{R}^{d_i}$  taking its values in a subset $\Sr_i$ 
of $\mathbb{R}^{m_i}$. 

Let us denote  for all $1 \leq  i \leq  \nbind$ by  $L_i$ the function defined on  $ \Sr_i \times \Theta$ by $L_i(s_i; \theta)\triangleq - \psi_i(\theta) + \pscal{s_i}{\phi_i(\theta)}$ and by $L: \Sr \times \Theta \to \Rset{}$ the function defined as $L(s,\theta)=\sum_i L_i(s_i; \theta)$ with $\Sr=\prod_i \Sr_i$ and $s=(s_1,\ldots,s_n)$.
For sake of simplicity, we omitted here all dependency on the observations $(y_i)_{1 \leq  i \leq  \nbind}$  since the considered stochasticity relies on the latent variables.

Finally let us denote by $(\gamma_k)_{k \geq 1}$ a sequence of positive step sizes. 

Moreover we assume that there exists a function
$\htheta : \ \Sr \rightarrow \Theta$, such that $
\forall s \in \Sr, \ \  \forall \theta \in \Theta, \ \ 
L(s; \htheta(s))\geq L(s; \theta).
$

\begin{itemize}\label{algo:SAEM}
	\item \textbf{Initialization step:}  Initialize arbitrarily for all $1 \leq i \leq \nbind $ $s_i^0$ and $\te_0$. 
	\item \textbf{Repeat until convergence the  three   steps defined at iteration $k$ by:}
	\begin{itemize}
		\item[$\circ$] \textbf{Simulation  step:}  for $1 \leq i \leq \nbind $ simulate a realization  $\MIS_i^k$ from  the conditional distribution given the observations $\OBS_i$ denoted by $p_i$ using  the current parameter value $\te_{k-1}$. 
		\item[$\circ$] \textbf{Stochastic approximation step:} compute the quantities for all  $1 \leq i \leq \nbind $
		\begin{equation*}
			\label{approsto}
			s_i^{k} = (1-\gamma_k)s_i^{k-1} +\gamma_k  S_i(\MIS_i^k) 
		\end{equation*}
		where $(\gamma_k)$ is a sequence of positive step sizes satisfying $\sum \gamma_k=\infty$ and $\sum \gamma_k^2 <~\infty$.
		\item[$\circ$] \textbf{Maximisation step:} update  of the parameter estimator  according to:
		\begin{eqnarray*}
			\te_{k}=  \argmax_{\te}  \sum_{i=1}^\nbind \left(   -\psi_i(\theta) + \pscal{s_i^k}{\phi_i(\theta)}   \right) = \hat{\theta}(s^{k})\\
		\end{eqnarray*}
	\end{itemize}
	\item \textbf{When convergence is reached, say  at iteration $K$ of the algorithm, evaluate the FIM estimator according to:}
	\begin{eqnarray*}
		\estiun^K &=& \frac{1}{n} \sum_{i=1}^n \hat{\Delta}_i\left(\hat{\theta}\left(s^{K}\right)\right) \hat{\Delta}_i\left(\hat{\theta}\left(s^{K}\right)\right)^t
	\end{eqnarray*}
	where $
	\hat{\Delta}_i(\hat{\theta}(s))  =  -\partial \psi_i(\hat{\theta}(s)) + \pscal{s_i}{\partial \phi_i(\hat{\theta}(s))}\
	$ for all $s$.
\end{itemize}

\begin{remark}
	In the cases where the latent variables can not be simulated from the conditional distribution, one can apply the extension coupling the stochastic algorithm with a Monte Carlo Markov Chain procedure as presented in \cite{Kuhn2004}. All the following results can be extended to this case.
\end{remark}

\subsubsection{Theoretical convergence properties}

In addition to the exponential family assumption for each individual likelihood, we also make the same type of regularity assumptions as those presented  in \cite{Delyon1999} at each individual level. These assumptions are detailed in the appendix section.

\begin{theorem}
	\label{theo:conv.algo}
	Assume that $(M1')$ and  $(M2')$,  $(M3)$ to  $(M5)$ and $(SAEM1)$ to $(SAEM4)$  are fulfilled. Assume also that with probability 1 $\mathrm{clos}(\{s_k\}_{k \geq 1})$ is a compact subset of $\Sr$. Let us define $\Lr=\{\theta \in\Theta, \Dt l(y;\theta)=0\}$  the set of stationary points of the observed log-likelihood $l$. Then, for all $\te_0 \in \Theta$, for fixed $n \in \mathbb{N}^*$, we get: $\lim_k d(\theta_k,\Lr)=0$ and  $\lim_k d(\estiun^k,\Ir)=0$ where $\Ir=\{I(\theta), \theta \in \Lr\}$.
\end{theorem}

\begin{proof}
	Let us denote by $S(Z)=(S_1(Z_1),\ldots,S_n(Z_n))$ the sufficient statistics of the model we consider in our approach. Note as recalled in \cite{Delyon1999}, these are not unique. Let us also define $H(Z,s)=S(Z)-s$ and $h(s)=\E_{\MIS|\OBS;\te}(S(Z))-s$.
	Assumptions $(M1')$ and $(M2')$  imply that assumptions  $(M1)$ and $(M2)$ of Theorem 5 of \cite{Delyon1999} are  fulfilled. Indeed these assumptions focus on expressions and regularity properties of the individual likelihood functions and the corresponding sufficient statistics for each index $i \in \{1,\ldots,n\}$.  Then by linearity of the log-likelihood function and of the stochastic approximation and applying Theorem 5 of \cite{Delyon1999}, we get that $\lim_k d(\theta_k,\Lr)=0$. 
	Moreover we get that for $1 \leq i \leq \nbind$, each sequence $(s_i^k)$ converges almost surely toward $\E_{\MIS_i|\OBS_i;\te} (S_i(\MIS_i) )$.
	Since  assumption $(M2')$ ensures that for all $1 \leq i \leq \nbind$ the functions $\psi_i$ and $\phi_i$ are twice continuously differentiable and assumption $(M5)$ ensures that the function $\hat{\theta}$ is continuously differentiable, the function $\Phi_n$ defined by $\Phi_n(s^{k})=\frac1{n}\sum_{i=1}^n \hat{\Delta}_i(\hat{\theta}(s^{k}))\hat{\Delta}_i(\hat{\theta}(s^{k}))$ is continuous. Therefore  we get that $\lim_k d(\estiun^k,\Ir)=0$.
\end{proof}

We now establish the asymptotic normality of the estimate $\vectestiun^k$ defined as $\vectestiun^k=\Phi_n(\bar{s}^{k})$ with $\bar{s}^{k}=\sum_{l=1}^k s^l /k$ using the results stated by \cite{Delyon1999}.  Let us denote by $Vect(A)$ the vector composed of the elements of the triangular superior part of matrix $A$ ordered by columns.

\begin{theorem}
	Assume that $(M1')$ and  $(M2')$,  $(M3)$ to  $(M5)$, $(SAEM1')$,  $(SAEM2)$,  $(SAEM3)$,  $(SAEM4')$ and $(LOC1)$ to $(LOC3)$ are fulfilled.  Then, there exists a regular stable stationary point $\theta^* \in \Theta$ such that $\lim_k \theta_k=\theta^*$ a.s. Moreover  the sequence $(\sqrt{k}(Vect(\vectestiun^k)-Vect(\vectestiun(\theta^*))))\mathds{1}_{\lim \| \theta_k-\theta^*\|=0 }$ converges in distribution toward a centered Gaussian random vector  when $k$ goes to infinity. The asymptotic covariance matrix is characterised.  
\end{theorem}

\begin{proof}
	The proof follows the lines of this of Theorem 7 of \cite{Delyon1999}. 
	Assumptions $(LOC1)$ to $(LOC3)$ are those of \cite{Delyon1999} and  ensure the existence of a regular stable stationary point $s^*$ for $h$ and therefore of $\theta^*=\hat{\theta}(s^*)$ for the observed log-likelihood $l$. Then applying Theorem 4 of 
	\cite{Delyon1999}, we get that:
	\begin{equation*}
		\sqrt{k}( \bar{s}^k - s^*) \mathds{1}_{\lim \| s^k-s^*\|=0 } \overset{\mathcal{L}}{ \rightarrow} \mathcal{N}(0, J(s^*)^{-1}  \Gamma(s^*) J(s^*)^{-1}  )\mathds{1}_{\lim \| s_k-s^*\|=0 }
	\end{equation*}
	where the function $\Gamma$ defined in assumption $(SAEM4')$ and $J$ is the Jacobian matrix of the function $h$. 
	Applying the Delta method, we get that:
	\begin{equation*}
		\sqrt{k}( Vect(\Phi_n(\bar{s}^k)) - Vect(\Phi_n(s^*))) \mathds{1}_{\lim \| s^k-s^*\|=0 } \overset{\mathcal{L}}{ \rightarrow} W\mathds{1}_{\lim \| s^k-s^*\|=0 }
	\end{equation*}
	where $W \sim 
	\mathcal{N}(0, \partial Vect(\Phi_n (s^*)) J(s^*)^{-1}  \Gamma(s^*) J(s^*)^{-1} \partial Vect(\Phi_n (s^*))^t )$
	which leads to the result.
\end{proof}

Note that as usually in stochastic approximation results, the rate $\sqrt{k}$ is achieved when considering an average estimator (see Theorem 7 of \cite{Delyon1999} e.g).

\subsubsection{Description of the algorithm for general latent variables models }
\label{subseq:generalmodel}
In  general settings, the SAEM algorithm can yet  be applied to approximate numerically the maximum likelihood estimate of the model parameter. Nevertheless  there are no more theoretical garantees of convergence for the algorithm. However we propose an extended version of our algorithm which allows to get an estimate  of the Fisher information matrix as a by-product of the estimation algorithm.

\begin{itemize}\label{algo:SAEMnonexpo}
	\item \textbf{Initialization step:}  Initialize arbitrarily $\Delta_i^0$ for all $1 \leq i \leq \nbind $, $Q_0$ and $\te_0$. 
	\item \textbf{Repeat until convergence the three steps defined at iteration $k$ by:}
	\begin{itemize}
		\item[$\circ$] \textbf{Simulation  step:}  for $1 \leq i \leq \nbind $ simulate a realization  $\MIS_i^k$ from  the conditional distribution given the observations $\OBS_i$, $p_i$, using the current parameter $\te_{k-1}$. 
		\item[$\circ$] \textbf{Stochastic approximation step:} compute the quantities for all  $1 \leq i \leq \nbind $
		\begin{eqnarray*}
			\label{approstononexpo}
			Q_{k}(\te)&=&(1-\gamma_k)Q_{k-1}(\te)+\gamma_k \sum_{i=1}^n \log f(\obs_i,\MIS_i^k;\te)\\
			\Delta_i^{k}&=&(1-\gamma_k)\Delta_i^{k-1} +\gamma_k \partial_\te \log \f(\obs_i,\MIS_i^k;\te_{k-1}) \nonumber
		\end{eqnarray*}
		\item[$\circ$] \textbf{Maximisation step:} update  of the parameter estimator  according to:
		\begin{eqnarray*}
			\te_{k}&=&  \argmax_{\te} Q_{k}(\te).
		\end{eqnarray*}
	\end{itemize}
	\item \textbf{When convergence is reached, say  at iteration $K$ of the algorithm, evaluate  the FIM estimator according to}:
	\begin{eqnarray*}
		\estiun^K &=& \frac{1}{n} \sum_{i=1}^n \Delta_i^K (\Delta_i^K )^t.
	\end{eqnarray*}
\end{itemize}

We illustrate through simulations  in a nonlinear mixed effects model the performance of this algorithm in Section \ref{sec:SimusNLMM}.

\section{Simulation study}
\label{sec:simul}

In this section, we investigate both the properties of the estimators $\estiun(\te)$ and $\estideux(\te)$ when the sample size $\nbind$ grows and the properties of the proposed algorithm when the number of iterations grows. 

\subsection{Asymptotic properties of the estimators $\estiun(\te)$ and $\estideux(\te)$}

\subsubsection{Simulation settings}
\label{sec:SimusLMM}

First  we consider the following linear mixed effects model $
\obs_{ij} = \beta + \mis_{i} + \varepsilon_{ij}, 
$
where $\obs_{ij} \in \R$ denotes the $j^{th}$ observation of individual $i$, $1\leq i \leq \nbind$, $1\leq j \leq \nobs$, $\mis_i \in \R$  the unobserved random effect of individual $i$ and $\varepsilon_{ij} \in \R$  the residual term. The random effects $(\mis_{i})$ are assumed independent and identically distributed such that $\mis_{i} \underset{i.i.d.}{\sim} \mathcal{N}(0,\eta^2)$, the residuals $(\varepsilon_{ij})$ are assumed independent and identically distributed such that $\varepsilon_{ij} \underset{i.i.d.}{\sim} \mathcal{N}(0,\sigma^2)$ and the sequences $(\mis_i)$ and $(\varepsilon_{ij})$ are assumed mutually independent. Here, the model parameters are given by $\te = (\beta, \eta^2, \sigma^2)$. We set $\beta=3$, $\eta^2=2$, $\sigma^2=5$ and $J=12$.

Second we consider the following Poisson mixture model where the distribution of each observation $\obs_i$ ($1\leq i \leq \nbind$) depends on a state variable $\mis_i$ which is latent leading to
$\obs_i|\mis_i=k  \sim  \mathcal{P}(\lambda_k)$ with $
P(\mis_i=k)  =  \alpha_k$ and 
$\sum_{k=1}^{K} \alpha_k  = 1. $
The model parameters are $\te=(\lambda_1,\ldots,\lambda_K,\alpha_1,\ldots,\alpha_{K-1})$. For the simulation study, we consider a mixture of $K=3$ components, and the following values for the parameters $\lambda_1=2$, $\lambda_2=5$, $\lambda_3=9$, $\alpha_1=0.3$ and $\alpha_2=0.5$.

\subsubsection{Results}

For each model, we generate $M=500$ datasets for different  sample sizes $\nbind \in \left\{20,100,500 \right\}$. We do not aim at estimating the model parameters. We assume them to be known, and in the following we denote by $\te^{\star}$ the true parameter value. For each value of $\nbind$ and for each $1 \leq m \leq M$, we derive $\estiun^{(m)}(\te^{\star})$ and $\estideux^{(m)}(\te^{\star})$. We compute the empirical bias and the root mean squared deviation of each component $(\ell,\ell')$ of the estimated matrix as:
$$
\frac{1}{M} \sum\limits_{m=1}^{M} I_{\nbind,sco,\ell,\ell'}^{(m)}(\te^\star) - \fisher_{\ell,\ell'}(\te^\star)  \; \; \;  \mathrm{and} \; \; \; \sqrt{\frac{1}{M} \sum\limits_{m=1}^{M} \left(I_{\nbind,sco,\ell,\ell'}^{(m)}(\te^\star) - \fisher_{\ell,\ell'}(\te^\star)\right)^2}.
$$

\begin{figure}[p]
	\centering
	\caption{\label{fig:AsymptDistLinMixt} Kernel density estimates of the normalized values $\sqrt{\nbind} \left(I_{n,sco,\ell,\ell'}^{(m)}(\te^\star) - \fisher_{\ell,\ell'}(\te^\star)\right)$ and $\sqrt{\nbind} \left(I_{n,obs,\ell,\ell'}^{(m)}(\te^\star) - \fisher_{\ell,\ell'}(\te^\star)\right)$ of some components of the estimated Fisher information matrix computed from the $M=500$ simulated datasets  when $\nbind=500$ (linear mixed effects model or the three graphs on the left, mixture model for the three graphs on the right).}
	\includegraphics[scale=0.35]{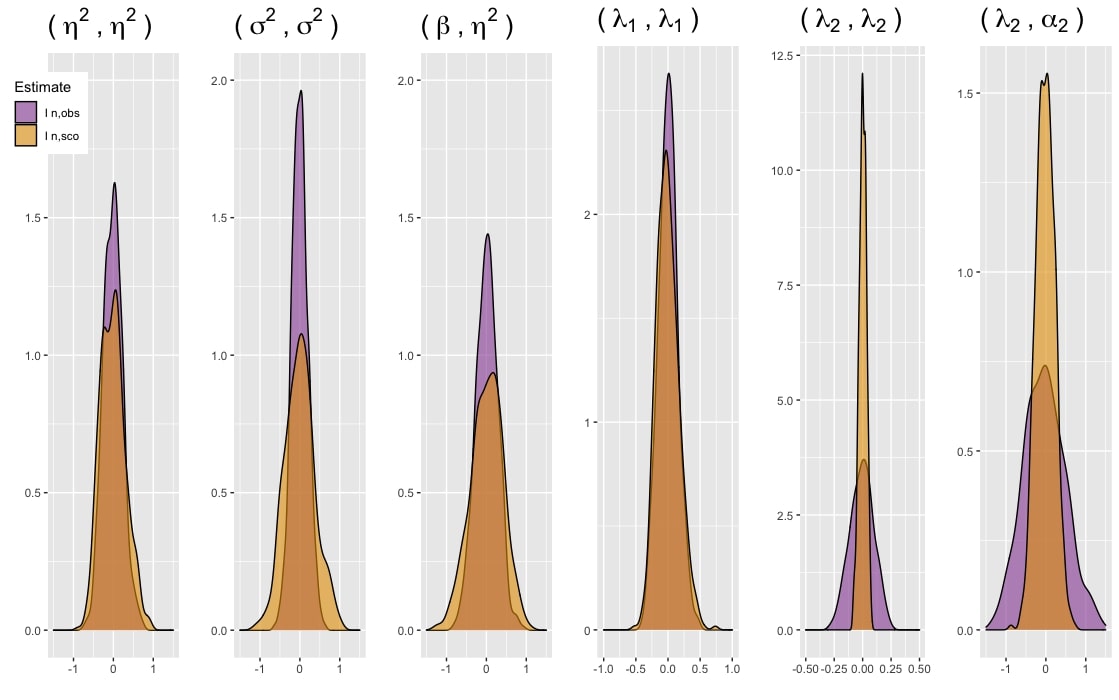}
\end{figure}

In the previous quantities, $\fisher(\te^\star)$ is explicit in the linear mixed effects model and  approximated by a Monte-Carlo estimation based on a large sample size in the Poisson mixture model. The results are presented in Tables \ref{tab:SimusLinVn} and \ref{tab:SimusLinWn} for the linear mixed effects model and in Tables \ref{tab:SimusMixtVn} and \ref{tab:SimusMixtWn} for the mixture model. We observe that whatever the model and whatever the components of $\estiun(\te^{\star})$ and $\estideux(\te^{\star})$, the bias is very small even for small values of $\nbind$. Note that in this particular model the second derivatives with respect to parameter $\beta$ is deterministic, which explains why the bias and the dispersion of the estimations $\estideux(\te^{\star})$ are zero for every value of $\nbind$. The bias and the standard deviation decrease as $\nbind$ increases overall, which illustrates the consistency of both M-estimators. We also represent in Figure \ref{fig:AsymptDistLinMixt} the distributions of the normalized estimations $\sqrt{\nbind} \left(\estiun^{(m)}(\te^\star) - \fisher(\te^\star)\right)$ and $\sqrt{\nbind} \left(\estideux^{(m)}(\te^\star) - \fisher(\te^\star)\right)$ for $\nbind=500$ for some components of the matrices. The empirical distributions have the shape of Gaussian distributions and illustrate the asymptotic normality of the two estimators. The numerical results highlight that neither $\estiun(\te^{\star})$ nor $\estideux(\te^{\star})$ is systematically better than the other one in terms of bias and asymptotic covariance matrix. In the same model, different behaviours can be observed depending on the components of the parameter vector.

\begin{table}[p]
	\caption{\label{tab:SimusLinVn} Linear mixed effects model. Empirical bias and squared deviation to the Fisher Information matrix (in brackets) of $\estiun$ for different values of $\nbind$.}
	\centering
	\begin{tabular}{|c|c|c|c|c|c|c|}
		\hline
		$\nbind$ & $\estiun(\beta,\beta)$ & $\estiun(\eta^2,\eta^2)$ & $\estiun(\sigma^2,\sigma^2)$ & $\estiun(\beta,\eta^2)$ & $\estiun(\beta,\sigma^2)$ & $\estiun(\eta^2,\sigma^2)$ \\
		\hline
		20 & 0.015& 0.007 & -0.007 & 0.002 &  -0.004 & -3.$10^{-4}$\\
		& (0.141) & (0.009) & (0.085) & (0.102) & (0.068) & (0.032)\\
		\hline
		100 & -0.001 & -2.$10^{-4}$ & -0.001 & 0.001 & 0.002 & 4.$10^{-4}$\\
		& (0.057) & (0.030) & (0.039) & (0.039) & (0.031) & (0.014)\\
		\hline
		500 & -0.001 & -7.$10^{-4}$ & -1.$10^{-4}$ & 5.$10^{-4}$ & -4.$10^{-4}$ & -4.$10^{-5}$ \\
		& (0.026) & (0.014) & (0.017) & (0.018) & (0.013) & (0.006)\\
		\hline
	\end{tabular}
\end{table}

\begin{table}[p]
	\caption{\label{tab:SimusLinWn} Linear mixed effects model. Empirical bias and squared deviation to the Fisher Information matrix (in brackets) of $\estideux$ for different values of $\nbind$.}
	\centering
	\begin{tabular}{|c|c|c|c|c|c|c|}
		\hline
		$\nbind$ & $\estideux(\beta,\beta)$ & $\estideux(\eta^2,\eta^2)$ & $\estideux(\sigma^2,\sigma^2)$ & $\estideux(\beta,\eta^2)$ & $\estideux(\beta,\sigma^2)$ & $\estideux(\eta^2,\sigma^2)$ \\
		\hline
		20 & 0.000 & 0.007 & -0.002 & 0.001 & 1.$10^{-4}$ & 5.$10^{-4}$\\
		& (0.000) & (0.058) & (0.042) & (0.058) & (0.005) & (0.005)\\
		\hline
		100 & 0.000 & -5.$10^{-4}$ & 2.$10^{-4}$ & -0.002 & -2.$10^{-4}$ & 4.$10^{-5}$ \\
		& (0.000) & (0.023) & (0.018) & (0.026) & (0.002) & (0.002)\\
		\hline
		500 & 0.000 & -5.$10^{-4}$ & -8.$10^{-5}$ & 4.$10^{-4}$ & 4.$10^{-5}$ & -4.$10^{-5}$\\
		& (0.000) & (0.011) & (0.009) & (0.012) & (0.001) & (0.001)\\
		\hline
	\end{tabular}
\end{table}

\begin{table}[p]
	\caption{\label{tab:SimusMixtVn} Mixture model. Empirical bias and squared deviation to the Fisher Information matrix (in brackets) of some components of $\estiun$ for different values of $\nbind$.}
	\centering
	\begin{tabular}{|c|c|c|c|c|c|c|}
		\hline
		$\nbind$  & $\estiun(\lambda_2,\lambda_2)$ & $\estiun(\lambda_3,\lambda_3)$ & $\estiun(\alpha_1,\alpha_1)$ & $\estiun(\alpha_2,\alpha_2)$ & $\estiun(\lambda_2,\lambda_3)$ & $\estiun(\lambda_3,\alpha_2)$ \\
		\hline
		20 &  8.$10^{-5}$ & 3.$10^{-5}$ & 0.060 & 0.047 & 9.$10^{-5}$ & -0.002 \\
		&  (0.007) & (0.015) & (1.202) & (1.056) & (0.003) & (0.110)  \\
		\hline
		100  & -3.$10^{-5}$ & -2.$10^{-4}$ & -0.040 & -0.041 & -7.$10^{-5}$ & 0.003 \\
		&  (0.003) & (0.007) & (0.526) & (0.469) & (0.001) & (0.046) \\
		\hline
		500  & 7.$10^{-5}$ & 7.$10^{-5}$ & 0.019 & 0.011 & 2.$10^{-5}$ & -1.$10^{-4}$ \\
		&  (0.001) & (0.003) & (0.232) & (0.205) & (0.001) & (0.021)  \\
		\hline
	\end{tabular}
\end{table}

\begin{table}[p]
	\caption{\label{tab:SimusMixtWn} Mixture model. Empirical bias and squared deviation to the Fisher Information matrix (in brackets) of some components of $\estideux$ for different values of $\nbind$.}
	\centering
	\begin{tabular}{|c|c|c|c|c|c|c|}
		\hline
		$\nbind$ &  $\estideux(\lambda_2,\lambda_2)$ & $\estideux(\lambda_3,\lambda_3)$ & $\estideux(\alpha_1,\alpha_1)$ & $\estideux(\alpha_2,\alpha_2)$ & $\estideux(\lambda_2,\lambda_3)$ & $\estideux(\lambda_3,\alpha_2)$ \\
		\hline
		20  & -3.$10^{-4}$ & 5.$10^{-4}$ & 0.060 & 0.047 & 9.$10^{-5}$ & 9.$10^{-4}$ \\
		& (0.022) & (0.009) & (1.202) & (1.055) & (0.003) & (0.034)  \\
		\hline
		100  & 2.$10^{-4}$ & -4.$10^{-4}$ & -0.040 & -0.041 & -7.$10^{-5}$ & -5.$10^{-4}$  \\
		&  (0.010) & (0.004) & (0.526) & (0.469) & (0.001) & (0.016)  \\
		\hline
		500  & -3.$10^{-4}$ & 1.$10^{-4}$ & 0.019 & 0.011 & 2.$10^{-5}$ & 5.$10^{-4}$ \\
		&  (0.005) & (0.002) & (0.232) & (0.205) & (6.$10^{-4}$) & (0.007)  \\
		\hline
	\end{tabular}
\end{table}

\subsection{Asymptotic properties of the stochastic approximation algorithm}
\label{sec:SimusNLMM}

\subsubsection{In curved exponential family model}

\begin{figure}[p]
	\centering
	\caption{\label{fig:BiaisNonLin} Non linear mixed effects model. Representation over iterations of the mean relative bias of the diagonal components of the estimated Fisher information matrix computed from the $M=500$ runs of the stochastic algorithm. Red line corresponds to $\estiun(\te)$ and blue line corresponds to $\estideux(\te)$. The burn-in iterations of the algorithm are not depicted.}
	\includegraphics[height=7cm]{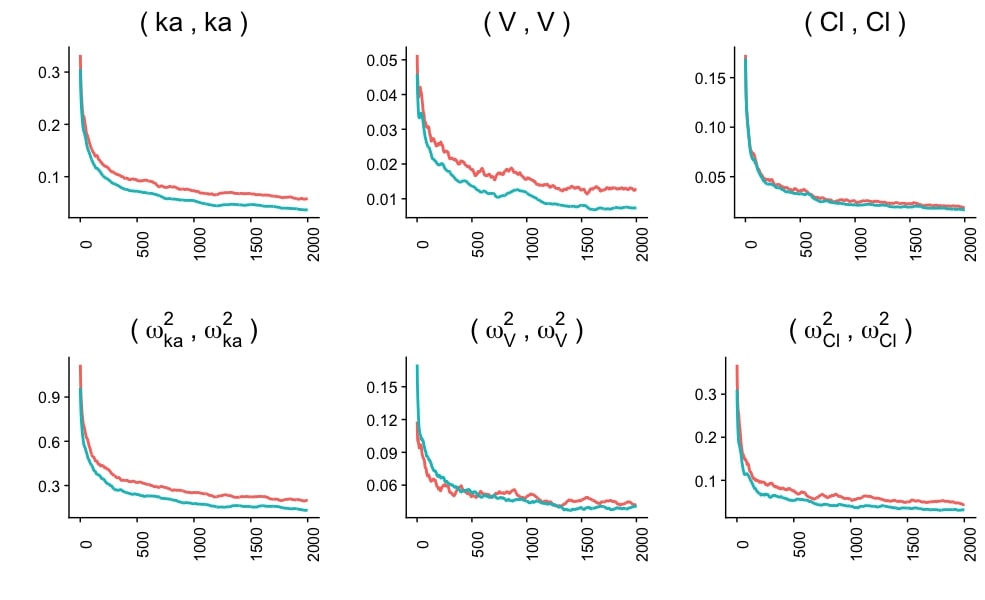}
\end{figure}

\begin{figure}[p]
	\centering
	\caption{\label{fig:SdNonLin} Non linear mixed effects model. Representation over iterations of the mean relative standard error of the diagonal components of the estimated Fisher information matrix computed from the $M=500$ runs of the stochastic algorithm. Red line corresponds to $\estiun(\te)$ and blue line corresponds to $\estideux(\te)$. The burn-in iterations of the algorithme are not depicted.}
	\includegraphics[height=7cm]{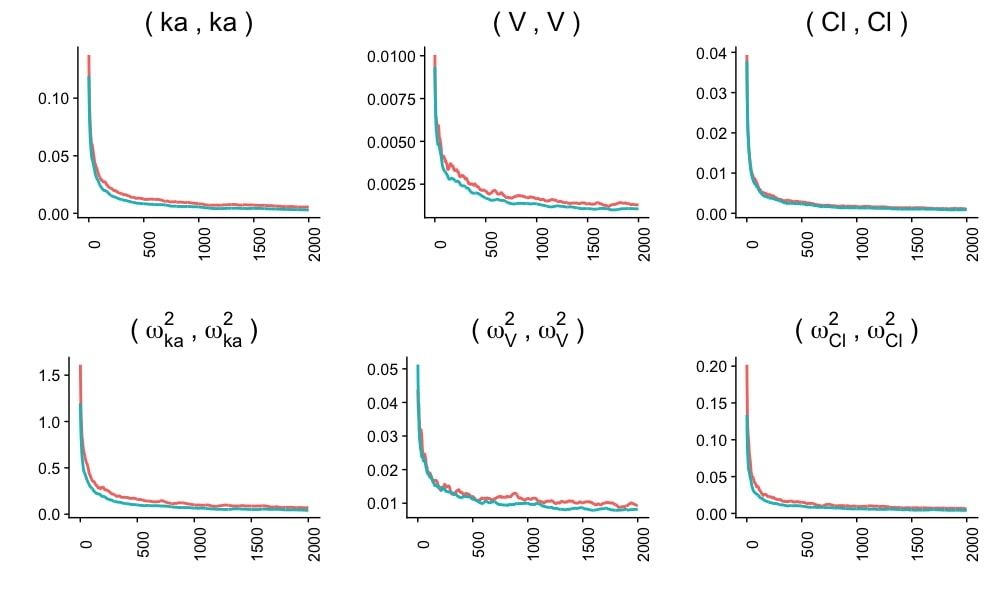}
\end{figure}

We consider the following nonlinear mixed effects model which is widely used in pharmacokinetics for describing the evolution of drug concentration over time:
\begin{equation}
	\obs_{ij}=\frac{d_i ka_{i}}{V_i ka_{i}-Cl_{i}}\left[e^{-\frac{Cl_{i}}{V_i} t_{ij}} - e^{-ka_{i} t_{ij}}\right] + \varepsilon_{ij},
	\label{eq:modelPK}
\end{equation}
where $ka_i$,  $Cl_i$ and $V_i$ are individual random parameters such that
$  \log ka_{i}  =  \log(ka) + \mis_{i,1}$, $
\log Cl_{i}  =  \log(Cl) + \mis_{i,2}$, 
$  \log V_i  =  \log(V) + \mis_{i,3}.$
For all $1 \leq i \leq n$, $1\leq j \leq \nobs$,
$\obs_{ij}$ denotes the measure of drug concentration on individual $i$  at time $t_{ij}$, $d_i$  the dose of drug administered to individual i, and $V_i$, $ka_i$ and $Cl_i$ respectively denote the volume of the central compartment, the drug's absorption rate constant and the drug's clearance of individual $i$. 
The terms $\mis_{i} = (\mis_{i,1},\mis_{i,2},\mis_{i,3})' \in \R^3$ are unobserved random effects which are assumed independent and identically distributed such that $\mis_i \underset{i.i.d.}{\sim} \mathcal{N}(0,\Omega)$, where $\Omega = \mathrm{diag}(\omega^2_{ka},\omega^2_{Cl},\omega^2_{V})$, the residuals $(\varepsilon_{ij})$ are assumed independent and identically distributed such that $\varepsilon_{ij} \underset{i.i.d.}{\sim} \mathcal{N}(0,\sigma^2)$ and the sequences $(\mis_i)$ and $(\varepsilon_{ij})$ are assumed mutually independent. Here, the model parameter is given by $\te = (ka,V,Cl,\omega^2_{ka},\omega^2_{V},\omega^2_{Cl},\sigma^2)$. In this model, as in a large majority of nonlinear mixed effects models, the likelihood does not have any analytical expression. As a consequence, neither the Fisher Information Matrix, nor the estimators $\estiun(\te)$, $\estideux(\te)$ have explicit expressions. However, as the complete data log-likelihood is explicit, stochastic approximations of $\estiun(\te)$, $\estideux(\te)$ can be implemented. We take the following values for the parameters $V=31$, $ka=1.6$, $Cl=2.8$, $\omega^2_V=0.40$, $\omega^2_{ka}=0.40$, $\omega^2_{Cl}=0.40$ and $\sigma^2=0.75$. We consider the same dose $d_i=320$ and the same observation times (in hours): $0.25$,$0.5$, $1$, $2$, $3.5$, $5$, $7$, $9$, $12$, $24$ for all the individuals. We simulate one dataset with $\nbind=100$ individuals under model \eqref{eq:modelPK}. On this simulated dataset, we run the stochastic approximation algorithm described in section \ref{algo:SAEM} for computing $\estiun(\hat{\te})$ together with $\hat{\te}$ and the algorithm of \cite{Delyon1999} for computing $\estideux(\hat{\te})$ $M=500$ times. We perform $K=3000$ iterations in total for each algorithm by setting $\gamma_k=0.95$ for $1 \leq k \leq 1000$ (burn in iterations) and $\gamma_k=(k-1000)^{-3/5}$ otherwise. At any iteration, we compute the empirical relative bias and the empirical relative standard deviation of each component $(\ell,\ell')$ of $I_{n,sco}$ defined respectively as: 
$$
\frac{1}{M} \sum\limits_{m=1}^{M} \frac{\widehat{I_{n,sco,\ell,\ell'}^{(k,m)}} - I_{n,sco,\ell,\ell'}^{\star}}{I_{n,sco,\ell,\ell'}^{\star}} \; \; \; \mathrm{and} 
\; \; \; \sqrt{\frac{1}{M} \sum\limits_{m=1}^{M} \left(\frac{\widehat{I_{n,sco,\ell,\ell'}^{(k,m)}} - I_{n,sco,\ell,\ell'}^{\star}}{I_{n,sco,\ell,\ell'}^{\star}}
	\right)^2} 
$$
where $\widehat{\estiun^{(k,m)}}$ denotes the estimated value of $\estiun(\hat{\te})$ at iteration $k$ of the $m^{th}$ algorithm. We compute the same quantities for $\estideux$. As the true values of $\estiun^{\star}=\estiun(\te^{\star})$ and $\estideux^{\star}=\estideux(\te^{\star})$ are not known, they are estimated by Monte-Carlo integration. The results are displayed in Figures \ref{fig:BiaisNonLin} and \ref{fig:SdNonLin}.

We observe that the bias and the standard deviations of the estimates of the components of both matrices decrease over iterations, and that for both estimates the bias is nearly zero when the convergence of the algorithm is reached.  According to these simulation results, there is no evidence that one method is better than the other in terms of bias or standard deviation.

\subsubsection{In general latent variable model}

\begin{figure}[p]
	\centering
	\includegraphics[scale=0.45]{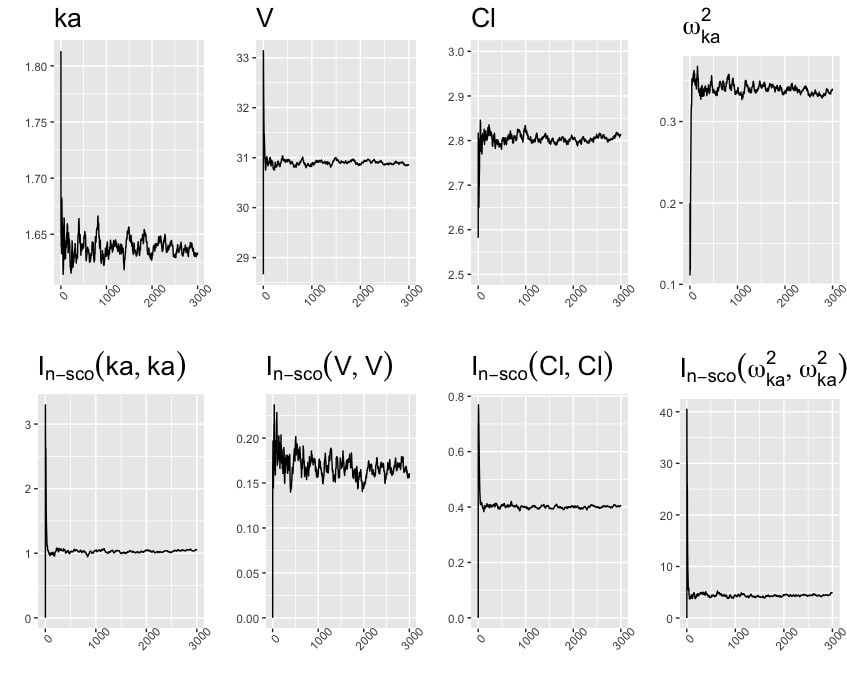}
	\caption{Convergence plot for some parameter estimates and for some diagonal components of $\estiun(\hat{\te})$ over iterations of the stochastic approximation algorithm.}
	\label{fig:convvn}
\end{figure}

We use model \eqref{eq:modelPK} again, but we now consider that individual parameter $V_i$ is fixed, {\it i.e.} $V_i \equiv V$ $\forall i = 1,\ldots,n$. The model is no longer exponential in the sense of equation \eqref{eq:curvedexpo}. We must therefore use the general version of the stochastic approximation algorithm from section \ref{subseq:generalmodel} to compute $\estiun(\hat{\te})$.  We simulate 500 datasets according to this model and we estimate $\estiun(\hat{\te})$ and $\hat{\te}$ for each one. We perform $K=3000$ iterations of the algorithm by setting $\gamma_k=k^{-0.501}$. We compute the 500 asymptotic confidence intervals of the model parameters $[\hat{\theta}^{(\ell)}_k - q_{1-\alpha/2} \, \hat{\sigma}^{(\ell)}_k , \hat{\theta}^{(\ell)}_k + q_{1-\alpha/2} \, \hat{\sigma}^{(\ell)}_k]$, $\ell =1,\ldots,6$ and then deduce from them empirical coverage rates. The $\hat{\sigma}^{(\ell)}_k$'s are obtained through the diagonal terms of the inversed $V_n(\hat{\theta}_k)$'s, and $q_{1-\alpha/2}$  stands for the quantile of order $1-\alpha/2$ of a standard Gaussian distribution with zero mean. We obtain for the six parameters $(ka,V,Cl,\omega^2_{ka},\omega^2_{Cl},\sigma^2)$ empirical covering rates of 0.946,0.928,0.962,0.944,0.950,0.942 respectively for a nominal covering rate of 0.95. This highlights that our estimate accurately quantifies the precisions of parameter estimates. Convergence graphs obtained from a simulated data set are shown in Figure \ref{fig:convvn}. Although theoretical guarantee is missing in non exponential models, the stochastic approximation algorithm proposed in section \ref{subseq:generalmodel} converges in practice on this example for both the estimation of the model parameters and the estimation of the Fisher information matrix. 

\subsection{Comparison with other methods}

To the best of our knowledge, although there exists contributions focusing on the estimation of the Fisher information matrix in latent variable models, there is currently no method based on the first derivatives of the log-likelihood. We compare to \cite{Meng2017} who proposed an iterative method based on numerical first order derivatives of the Q function that is computed at each E-step of the EM algorithm. The model used by \cite{Meng2017} in their simulation study is a mixture of two Gaussian distributions with unknown expectations $\mu_1$ and $\mu_2$, fixed variances equal to $1$ and unknown proportion $\pi$. The model parameters are denoted by $\theta=(\mu_1,\mu_2,\pi)$.

We simulate 10000 datasets according to this Gaussian mixture model, using the same setting as \cite{Meng2017}, {\it i.e.} $n=750$, $\pi=2/3$, $\mu_1=3$ and $\mu_2=0$. For each dataset $k=1,\ldots,10000$, we compute the parameter maximum likelihood estimate $\hat{\theta}_k = (\hat{\pi}_k,\widehat{\mu_1}_k,\widehat{\mu_2}_k)$ with an EM algorithm and then we derive $\estiun(\hat{\theta}_k)$ directly according to formula \eqref{eq:vnmis} contrary to \cite{Meng2017} who used an iterative method. We compute the empirical mean of the  10000 estimated matrices leading to: 
$$
\frac1{10000} \sum _k \estiun(\hat{\theta}_k)= \begin{pmatrix}
	2685.184 & -211.068 & -251.808\\
	-211.068 & 170.927 & -61.578 \\
	-251.808 & -61.578 & 392.859\\
\end{pmatrix}.
$$
Comparison with the results of \cite{Meng2017} is delicate since their numerical illustration of their method is based on a single simulated dataset thus potentially sensitive to sampling variations. However, they provide an estimation of the Fisher information matrix from this unique dataset   
$$
I_{Meng} = \begin{pmatrix}
	2591.3 & -237.9 & -231.8\\
	-237.9 & 155.8 & -86.7\\
	-231.8 & -86.7 & 394.5
\end{pmatrix}.
$$
Our results are coherent with their ones. To check the reliability of our results, we then compute as above the 10000 asymptotic confidence intervals of the three model parameters. We obtain for the three parameters $(\pi,\mu_1,\mu_2)$ empirical covering rates of $0.953,0.949,0.951$ respectively for a nominal covering rate of $0.95$. Thus $\estiun$  accurately quantifies the precisions of parameter estimates.

\section{Conclusion and discussion}

	In this work, we address the estimation of the Fisher information matrix in general latent variable models. We focus on the empirical Fisher information matrix which is a moment estimate of the covariance matrix of the score.  We propose a stochastic approximation algorithm to compute this estimate when it can not be calculated analytically and establish its theoretical convergence properties. We carry out a simulation study in mixed effects model and in a Poisson mixture model to compare the performances of several estimates, namely the considered empirical Fisher information matrix and the observed  Fisher information matrix. We emphasize that the  empirical FIM requires less regularity assumptions than the  observed  FIM. From a computational point of view, the implementation of the algorithm for evaluating the empirical FIM only involves the first derivatives of the log-likelihood, in contrary to the one for evaluating the observed FIM which involves the second derivatives of the log-likelihood.

The main perspective of this work is to adapt the procedure for statistical models whose derivatives of the log-likelihood have no tractable expressions, coupling the algorithm with numerical derivative procedures.

\section{Appendix}
It is assumed that the random variables $s^0, z_1, z_2, \cdots$ are 
defined on the same probability space $(\Omega, \mathcal{A}, P)$. We denote $\Fr = \{ \Fr_k \}_{k \geq 0}$ the increasing family of 
$\sigma$-algebras generated by the random variables $s_0, z_1, z_2, \cdots, z_k$. We assume the following conditions:

\begin{itemize}
	\item {\bf (M1')} The parameter space $\Theta$ is an open subset of $\mathbb{R}^{p}$. The individual complete data likelihood function is given for all $i=1,\ldots,\nbind$  by:
	\begin{equation*} 
		f_i(z_i;\theta)
		= \exp\left(-\psi_i(\theta) + \pscal{S_i(z_i)}{\phi_i(\theta)}\right),
	\end{equation*}
	where  $\pscal{\cdot}{\cdot}$ denotes the scalar product, $S_i$ is a Borel
	function on $\mathbb{R}^{d_i}$  taking its values in an open subset $\Sr_i$ 
	of $\mathbb{R}^{m_i}$. Moreover, the convex hull of $S(\mathbb{R}^{\sum d_i})$  is included in $\Sr$ and for all $\theta \in \Theta$ $\int S(z) \prod p_i(z_i;\theta) \mu(dz) < \infty$
	\item {\bf (M2')} Define for each $i$ $L_i : \Sr_i \times \Theta \to \Rset{}$ as $ 
	L_i(s_i; \theta)\triangleq - \psi_i(\theta) + \pscal{s_i}{\phi_i(\theta)}.
	$
	The functions $\psi_i$ and $\phi_i$ are twice 
	continuously differentiable on $\Theta$. 
	\item {\bf (M3)} 
	The function $\bar{s} : \Theta \rightarrow \Sr$ defined as
	$
	\bar{s}(\theta) \triangleq \int S(z) p(z; \theta) \mu(dz)
	$
	is continuously differentiable on $\Theta$.
	\item {\bf (M4)} The function $l : \Theta \rightarrow \Rset{}$ defined as 
	$
	l(\theta) \triangleq \log g(\theta) = \log \int_{\mathbb{R}^{d_z}} f(z;\theta) \mu(dz)
	$
	is continuously differentiable on $\Theta$ and
	$
	\Dt \int f(z; \theta) \mu(dz)= \int \Dt f(z; \theta)  \mu(dz).
	$
	\item {\bf (M5)} There exists a continuously differentiable function
	$\htheta : \ \Sr \rightarrow \Theta$, such that:
	\begin{equation*} 
		\forall s \in \Sr, \ \  \forall \theta \in \Theta, \ \ 
		L(s; \htheta(s))\geq L(s; \theta).
	\end{equation*}  
\end{itemize}

In addition, we define:
\begin{itemize}
	\item {\bf (SAEM1)} For all $k$ in $\Nset$, $\gamma_k \in [0,1]$, 
	$\sum_{k=1}^\infty \gamma_k = \infty$ and  $\sum_{k=1}^\infty \gamma_k^2 < 
	\infty$.
	\item  {\bf (SAEM2)}  $l  : \Theta  \rightarrow  \Rset{}$ and  $\htheta :  \Sr
	\rightarrow \Theta$ are $m$ times differentiable, where $m$ is the integer such that $\Sr$ is an open subset of $\Rset{m}$. 
	\item {\bf (SAEM3)}
	For all positive Borel functions $\Phi$ $E[ \Phi( z_{k+1}) | \mathcal{F}_k ] = 
	\int \Phi( z  ) p ( z; \theta_k) \mu( dz).$
	\item {\bf (SAEM4)} For all $\theta \in \Theta$, $ \int \| S(z) \|^2 p( z; \theta) \mu (d z) < \infty$, and the function 
	\begin{equation*}
		\begin{split}
			\Gamma(\theta) \triangleq \mathrm{Cov}_\theta [S(z)]\triangleq &\int
			S(z)^t S(z) p(z;\theta)\mu(dz)\\
			&-\left[\int S(z)p(z;\theta)\mu(dz)\right]^t\left[\int S(z)p(z;\theta)\mu(dz)\right]
		\end{split}
	\end{equation*}
	is continuous w.r.t. $\theta$.
\end{itemize}
We also define assumptions required for the normality result:
\begin{itemize}
	\item{\bf (SAEM1')} For all $k$ in $\Nset$, $\gamma_k \in [0,1]$, 
	$\sum_{k=1}^\infty \gamma_k = \infty$ and  $\sum_{k=1}^\infty \gamma_k^2 < 
	\infty$. There exists $\gamma^*$ such that $\lim k^\alpha /\gamma_k =\gamma^*$, and $\gamma_k / \gamma_{k+1}=1 + O(k^{-1})$.
	\item {\bf (SAEM4')} For some $\alpha>0$, $\sup_\theta \E_\theta(\|S(Z)\|^{2+\alpha})< \infty$ and $\Gamma$ is continuous w.r.t.~$\theta$.
	\item {\bf (LOC1)} The stationary points of $l$ are isolated: any compact subset of $\Theta$ contains only a finite number of such points.
	\item {\bf (LOC2)} For every stationary point $\theta^*$, the matrices $\E_\theta^* (\partial_\theta L(S(Z),\theta^*) (\partial_\theta L(S(Z),\theta^*))^t) $  and $ \partial_\theta^2  L(\E_\theta^* (S(Z)),\theta^*)$ are positive definite.
	\item {\bf (LOC3)} The minimum eigenvalue of the covariance matrix
	$R(\theta)=\E_\theta((S(Z)-\bar{s}(\theta))(S(Z)-\bar{s}(\theta))^t)$
	is bounded away from zero for $\theta$ in any compact subset of $\Theta$.
\end{itemize}

\def\cprime{$'$} \def\cprime{$'$}


\begin{thebibliography}{8}
\addcontentsline{toc}{section}{References}


\expandafter\ifx\csname natexlab\endcsname\relax\def\natexlab#1{#1}\fi
\providecommand{\bibinfo}[2]{#2}
\ifx\xfnm\relax \def\xfnm[#1]{\unskip,\space#1}\fi
\bibitem[1]{VanderVaart2000}
\bibinfo{author}{Van der Vaart, A. W.} \bibinfo{year}{(2000)} \bibinfo{title}{Asymptotic Statistics}, \bibinfo{series}{Cambridge Series in Statistical and Probabilistic Mathematics}, \bibinfo{publisher}{Cambridge University Press}.
\bibitem[2]{charkhi2018asymptotic}
\bibinfo{author}{Charkhi, A. and Claeskens, G.} \bibinfo{year}{(2018)}
\newblock \bibinfo{title}{Asymptotic post-selection inference for the Akaike information criterion},
\newblock \bibinfo{journal}{Biometrika}, \bibinfo{volume}{{\bf 105}}\bibinfo{issue}{(3)}, \bibinfo{pages}{645--664}. \bibinfo{doi}{\url{https://doi.org/10.1093/biomet/asy018}}
\bibitem[3]{baey2019asymptotic}
\bibinfo{author}{Baey, C., Courn{\`e}de, P.-H. and Kuhn, E.} \bibinfo{year}{(2019)}
\newblock \bibinfo{title}{Asymptotic distribution of likelihood ratio test statistics for variance components in nonlinear mixed effects models},
\newblock \bibinfo{journal}{Computational Statistics \& Data Analysis}, \bibinfo{volume}{{\bf 135}}, \bibinfo{pages}{107--122}.
\bibinfo{doi}{\url{https://doi.org/10.1016/j.csda.2019.01.014}}
\bibitem[4]{le2021fisher}
\bibinfo{author}{Le Brigant, A., Preston, S. C. and Puechmorel, S.} \bibinfo{year}{(2021)}
\newblock \bibinfo{title}{Fisher-Rao geometry of Dirichlet distributions},
\newblock\bibinfo{journal}{Differential Geometry and its Applications}, \bibinfo{volume}{{\bf 74}}, 101702. \bibinfo{doi}{\url{https://doi.org/10.1016/j.difgeo.2020.101702}} 
\bibitem[5]{Technow2015}
\bibinfo{author}{Technow, F., Messina, C. D., Totir, L. R. and Cooper, M.} \bibinfo{year}{(2015)}
\newblock \bibinfo{title}{Integrating crop growth models with whole genome prediction through approximate Bayesian computation},
\newblock \bibinfo{journal}{PloS one},\bibinfo{volume}{{\bf 10}}\bibinfo{issue}{(6)}, e0130855.
\bibinfo{doi}{\url{https://doi.org/10.1371/journal.pone.0130855}}
\bibitem[6]{Picard2007}
\bibinfo{author}{Picard, F., Robin, S., Lebarbier, E. and Daudin, J.-J.} \bibinfo{year}{(2007)}
\newblock \bibinfo{title}{A Segmentation/Clustering Model for the Analysis of Array CGH Data},
\newblock \bibinfo{journal}{Biometrics}, \bibinfo{volume}{{\bf 63}}, \bibinfo{pages}{758--766}.
\bibinfo{doi}{\url{https://doi.org/10.1111/j.1541-0420.2006.00729.x}}
\bibitem[7]{Gloaguen2014}
\bibinfo{author}{Gloaguen, P., Mah\'evas, S., Rivot, E., Woillez, M., Guitton, J., Vermard, Y. and Etienne, M.-P.} \bibinfo{year}{(2014)}
\newblock \bibinfo{title}{An autoregressive model to describe fishing vessel movement and activity},
\newblock \bibinfo{journal}{Environmetrics}, \bibinfo{volume}{{\bf 26}}(\bibinfo{issue}{1}), \bibinfo{pages}{17--28}.
\bibinfo{doi}{\url{https://doi.org/10.1002/env.2319}}
\bibitem[8]{mixfim2018}
\bibinfo{author}{Riviere-Jourdan, M.-K. and Mentre, F.} \bibinfo{year}{(2018)}
\newblock \bibinfo{title}{MIXFIM: Evaluation of the FIM in NLMEMs using MCMC}. \url{https://CRAN.R-project.org/package=MIXFIM}
\bibitem[9]{efron1978assessing}
\bibinfo{author}{Efron, B. and Hinkley, D. V.} \bibinfo{year}{(1978)}
\newblock \bibinfo{title}{Assessing the accuracy of the maximum likelihood estimator: Observed versus expected Fisher information},
\newblock \bibinfo{journal}{Biometrika}, \bibinfo{volume}{{\bf 65}}(\bibinfo{issue}{3}), \bibinfo{pages}{457--483}.
\bibinfo{doi}{\url{https://doi.org/10.2307/2335893}}
\bibitem[10]{Woodbury1972}
\bibinfo{author}{Orchard, T. and Woodbury, M. A.} \bibinfo{year}{(1972)}
\newblock \bibinfo{title}{A missing information principle: theory and applications},
\newblock \bibinfo{booktitle}{Proceedings of the Berkeley Symposium on Mathematical Statistics and Probability}.
\bibitem[11]{McLachlan2008}
\bibinfo{author}{McLachlan, G.-J. and Krishnan, T.} \bibinfo{year}{(2008)} \bibinfo{title}{The EM algorithm and extensions}, \bibinfo{series}{Wiley series in probability and statistics}, \bibinfo{publisher}{Wiley}.
\bibitem[12]{Louis1982}
\bibinfo{author}{Louis, T. A.} \bibinfo{year}{(1982)}
\newblock \bibinfo{title}{Finding the Observed Information Matrix when Using the EM Algorithm},
\newblock \bibinfo{journal}{Journal of the Royal Statistical Society. Series B (Methodological)}, \bibinfo{volume}{{\bf 44}}(\bibinfo{issue}{2}), \bibinfo{pages}{226--233}.
\bibinfo{doi}{\url{https://doi.org/10.1111/j.2517-6161.1982.tb01203.x}}
\bibitem[13]{Delyon1999}
\bibinfo{author}{Delyon, B. and Lavielle, M. and Moulines, E.} \bibinfo{year}{(1999)}
\newblock \bibinfo{title}{Convergence of a stochastic approximation version of the EM algorithm},
\newblock \bibinfo{journal}{Annals of Statistics}, \bibinfo{volume}{{\bf 27}}(\bibinfo{issue}{1}), \bibinfo{pages}{94--128}.
\bibitem[14]{Oakes1999}
\bibinfo{author}{Oakes, D.} \bibinfo{year}{(1999)}
\newblock \bibinfo{title}{Direct calculation of the information matrix via  the EM algorithm},
\newblock \bibinfo{journal}{Journal of the Royal Statistical Society. Series B (Methodological)}, \bibinfo{volume}{{\bf 61}}(\bibinfo{issue}{2}), \bibinfo{pages}{479--482}.
\bibinfo{doi}{\url{ https://doi.org/10.1111/1467-9868.00188}}
\bibitem[15]{Meng1991}
\bibinfo{author}{Meng, X.-L. and Rubin, D. B.} \bibinfo{year}{(1991)}
\newblock \bibinfo{title}{Using EM to obtain asymptotic variance-covariance matrices: the SEM algorithm},
\newblock \bibinfo{journal}{Journal of the American Statistical Association}, \bibinfo{volume}{{\bf 86}}(\bibinfo{issue}{416}), \bibinfo{pages}{899--909}.
\bibinfo{doi}{\url{ https://doi.org/10.1080/01621459.1991.10475130}}
\bibitem[16]{Meng2017}
\bibinfo{author}{Meng, L. and Spall, J. C.} \bibinfo{year}{(2017)}
\newblock \bibinfo{title}{Efficient computation of the Fisher information matrix in the EM algorithm},
\newblock \bibinfo{booktitle}{Annual Conference on Information Sciences and Systems (CISS)}.
\bibitem[17]{Scott2002}
\bibinfo{author}{Scott, W.-A.} \bibinfo{year}{(2002)}
\newblock \bibinfo{title}{Maximum likelihood estimation using the empirical fisher information matrix},
\newblock \bibinfo{journal}{Journal of Statistical Computation and Simulation}, \bibinfo{volume}{{\bf 72}}(\bibinfo{issue}{8}), \bibinfo{pages}{599--611}.
\bibinfo{doi}{\url{https://doi.org/10.1080/00949650213744}}
\bibitem[18]{kunstner2019limitations}
\bibinfo{author}{Kunstner, F., Hennig, P. and Balles, L.} \bibinfo{year}{(2019)}
\newblock \bibinfo{title}{Limitations of the empirical fisher approximation for natural gradient descent},
\newblock \bibinfo{booktitle}{Advances in neural information processing systems}, \bibinfo{volume}{32}.
\bibitem[19]{Nie2006}
\bibinfo{author}{Nie, L.} \bibinfo{year}{(2006)}
\newblock \bibinfo{title}{Strong consistency of the maximum likelihood estimator in generalized linear and nonlinear mixed-effects models},
\newblock \bibinfo{journal}{Metrika}, \bibinfo{volume}{{\bf 63}}(\bibinfo{issue}{2}), \bibinfo{pages}{123--143}.
\bibinfo{doi}{\url{ https://doi.org/10.1007/s00184-005-0001-3}}
\bibitem[20]{Silvapulle2011}
\bibinfo{author}{Silvapulle, M. J. and Sen, P. K.} \bibinfo{year}{(2011)} \bibinfo{title}{Constrained statistical inference: Order, inequality, and shape constraints}, \bibinfo{volume}{912}, \bibinfo{publisher}{John Wiley \& Sons}.
\bibitem[21]{fisher1925}
\bibinfo{author}{Fisher, R.A.} \bibinfo{year}{(1925)} \bibinfo{title}{Statistical methods for research workers}, \bibinfo{publisher}{John Wiley \& Sons}.
\bibitem[22]{Kuhn2004}
\bibinfo{author}{Kuhn, E. and Lavielle, M.} \bibinfo{year}{(2004)}
\newblock \bibinfo{title}{Coupling a stochastic approximation version of EM with an MCMC procedure},
\newblock \bibinfo{journal}{ESAIM P\&S}, \bibinfo{volume}{{\bf 8}}, \bibinfo{pages}{115--131}.
\bibinfo{doi}{\url{https://doi.org/10.1051/ps:2004007}}
\bibitem[23]{jaffrezic2006genetic}
\bibinfo{author}{Jaffr{\'e}zic, F., Meza, C., Lavielle, M. and Foulley, J.-L.} \bibinfo{year}{(2006)}
\newblock \bibinfo{title}{Genetic analysis of growth curves using the SAEM algorithm},
\newblock \bibinfo{journal}{Genetics Selection Evolution}, \bibinfo{volume}{{\bf 38}}(\bibinfo{issue}{6}), \bibinfo{pages}{583--600}.
\bibinfo{doi}{\url{https://doi.org/10.1051/gse:2006023}}




\end{thebibliography}
\end{document}